# Two distinct sequences of blue straggler stars in the globular cluster M30

F.R. Ferraro[1], G. Beccari[2], E. Dalessandro[1], B. Lanzoni[1], A. Sills[3], R. T. Rood[4], F. Fusi Pecci[5], A.I. Karakas[6], P. Miocchi[1] & S. Bovinelli[1]

[1]*Department of Astronomy, University of Bologna, Via Ranzani, 1, 40127 Bologna, Italy*

[2]*ESA, Space Science Department, Keplerlaan 1, 2200 AG Noordwijk, Netherlands*

[3]*McMaster Univ., Dep. of Physics and Astronomy, 1280 Main Street West, Hamilton, Canada*

[4]*Astronomy Department, University of Virginia, P.O. Box 400325, Charlottesville, VA, 22904*

[5]*INAF, Osservatorio Astronomico di Bologna, via Ranzani 1, 40127 Bologna, Italy*

[6]*Research School of Astronomy and Astrophysics, Mt Stromlo Observatory, Weston Creek ACT 2611, Australia*

**Stars in globular clusters are generally believed to have all formed at the same time, early in the Galaxy's history[1]. 'Blue stragglers' are stars massive enough[2] that they should have evolved into white dwarfs long ago. Two possible mechanisms have been proposed for their formation: mass transfer between binary companions[3] and stellar mergers resulting from direct collisions between two stars[4]. Recently, the binary explanation was claimed to be dominant[5]. Here we report that there are two distinct parallel sequences of blue stragglers in M30. This globular cluster is thought to have undergone 'core collapse', during which both the collision rate and the mass transfer activity in binary systems would have been**

**enhanced[6]. We suggest that the two observed sequences arise from the cluster core collapse, with the bluer population arising from direct stellar collisions and the redder one arising from the evolution of close binaries that are probably still experiencing an active phase of mass transfer.**

To investigate the blue straggler star (BSS) content in M 30, we used a time-series of 44 high-resolution images obtained with the NASA Hubble Space Telescope (Supplementary Information). The colour-magnitude diagram (CMD) derived by combining these measures has revealed the existence of two well-separated and almost parallel sequences of BSSs (hereafter 'red BSSs' and 'blue BSSs'; Fig. 1). The two sequences are similarly populated, consisting of 21 and 24 stars, respectively.

The detected BSSs are substantially more concentrated towards the cluster centre than 'normal' cluster stars, either along the subgiant branch or the horizontal branch (Fig. 2a). According to a Kolmogorov-Smirnov test, the probability that the BSSs and the subgiant- or horizontal-branch stars are drawn from the same distribution is only $\sim 10^{-3}$ (that is, they differ at a significance level of more than $4\sigma$). This result confirms that BSSs are more massive[2] than the majority of the cluster stars and that mass segregation has been active in this cluster. Moreover, when we consider the distribution of the two BSS subpopulations separately, we find that the red BSSs are more centrally segregated than the blue BSSs. Indeed, no red BSSs are observed at an angular distance of $r$>30 arcsec (corresponding to ~ 1.3 pc; see Supplementary Table 1) from the cluster centre (see Fig. 2a). Even though in this case the level of significance ($1.5\sigma$) is marginal because of the small number of objects, this evidence is suggestive of different formation histories for BSSs belonging to the two sequences. Furthermore, whereas the radial distribution of BSSs in many clusters is found to be bimodal[7] (with a dominant peak at the centre, a dip at intermediate radii and a rising branch in the outer regions), in the case of M 30 there is no evidence of an increase at large distances from the centre: more than 80% of the entire BSS population is confined within the inner 100 arcsec (~4.2 pc), and the radial distribution then stays nearly constant for greater radii. This suggests that dynamical friction has already affected a large portion of the cluster, so that almost the entire population of BSS has sunk into the centre[8].

There is further evidence that M 30 is a highly evolved cluster from a dynamical point of view. The density profile has a steep power-law cusp in the central 5-6 arcsec (~0.2 pc, Fig. 3), suggesting that M 30 already experienced the core collapse (Supplementary Information)[9]. M 30's dynamically evolved state, combined with several suggestions[4,7,10,11] that cluster dynamics and BSS formation processes could be linked, indicates that the dual BSS sequence is probably connected to the cluster's dynamical history. In particular, during core collapse the central density rapidly increases, bringing a concomitant increase of gravitational interactions[6] able to trigger the formation of new BSSs through both direct stellar collisions and enhanced mass transfer activity in dynamically shrunk binary systems. When considering the entire population of detected BSSs (population size, $N_{BSS}$=45) and HB stars ($N_{HB}$=90), the BSS specific frequency, $F^{BSS}=N_{BSS}/N_{HB}$, is equal to 0.5, a value not particularly high in comparison with that of other clusters[12]. However, the value of $F^{BSS}$ varies significantly over the surveyed area, reaching the value of ~1.55 when only the central cusp of the star density profile (5-6 arcsec) is considered (Fig. 2b). So far, this is the highest value ever measured for the BSS specific frequency in any globular cluster, and it further supports the possibility that in M 30 we are observing the effect of an enhanced gravitational interaction activity on single and binary stars.

To investigate this possibility, we have compared the observations with the predictions of evolutionary models of BSSs formed by direct collisions between two main-sequence stars[13] with metallicities of $Z = 10^{-4}$ and masses ranging between 0.4 $M_{\odot}$ and 0.8 $M_\odot$ (thus producing BSSs with masses between 0.8 and 1.6 $M_\odot$). As shown in Fig. 4, the blue BSS sequence is well fitted by collisional isochrones corresponding to ages of 1-2 Gyr, with the brightest blue BSS being slightly less luminous than the collision product of two turn-off-mass stars (0.8 $M_\odot$+0.8 $M_\odot$). The observed number of stars in the blue BSS sequence is in good agreement with the expected[14] number of BSSs formed by direct collisions during the last 1-2 Gyr in a cluster with total absolute magnitude comparable to that of M 30 (Supplementary Table 1). Hence, we conjecture that 1-2 Gyr ago some dynamical process (possibly core collapse) produced the BSS population that is now observable along the blue BSS sequence. The origin of the red BSSs should be different, as this sequence is too red to be properly reproduced by collisional isochrones corresponding to any age. Figure 4 also shows the location in the CMD of single-star isochrones[15] computed for a metallicity $Z = 2\times10^{-4}$ and shifted by the distance modulus and reddening of M 30[16]. The 13 Gyr single-star isochrone fits the main cluster evolutionary sequences quite well, whereas the BSS sequences are significantly offset with respect to the 0.5 Gyr single-star isochrone, which can be adopted as the cluster ZAMS. Of particular note is the fact that the red BSS sequence is ~0.75 mag brighter than the reference ZAMS. According to the results of recent binary-evolution models[17], during the mass-transfer phase (which can last several gigayears, that is, a significant fraction of the binary evolution timescale), a population of binary systems defines a low-luminosity boundary ~0.75 mag above the ZAMS in the BSS region (see fig. 5 of ref. 17). Hence, the BSSs that we observe along the red BSS sequence could be binary systems still experiencing an active phase of mass-exchange.

As result of normal stellar evolution, in a few gigayears both collisional and mass-transfer products will populate the region between the two sequences. The fact that we currently see two well-separated sequences supports the hypothesis that both the blue and the red BSS sequences were generated by a recent and short-lived event, instead of a continuous formation process. A picture in which the two BSS sequences were generated by the same dynamical event is therefore emerging. As suggested by the shape of the density profile and by the location of the blue BSSs in the CMD, 1-2 Gyr ago M 30 may have undergone core collapse. This is known to increase the gravitational interaction rate significantly and it may therefore have boosted the formation of BSSs: the blue BSSs arise from direct stellar collisions, while the red BSSs are the result of the evolution of binary systems which first sank into the cluster centre because of the dynamical friction (or they were already present into the cluster core), and were then brought into the mass-transfer regime by hardening processes induced by gravitational interactions during core collapse. The detected double BSS sequence could be the signature imprinted onto a stellar population by core collapse, with the red BSS sequence being the outcome of the "binary-burning" process expected to occur in the cluster core during the late stages of the collapse[18,19]. The proposed picture leads to a testable observational prediction: the red BSS sequence should be populated by binaries with short orbital periods.

A recent paper[5] suggested that the dominant BSS formation channel is the evolution of binary systems, independent of the dynamical state of the parent cluster. Our discovery shows that binary evolution alone does not paint a complete picture: dynamical processes can indeed play a major role in the formation of BSSs. An appropriate survey of the central regions of other core-collapsed clusters would help to clarify whether the double BSS sequence is a common signature of the core-collapse phenomenon. Moreover, detailed spectroscopic investigations are worth performing to obtain a complete characterization of the BSS properties (orbital periods, rotation velocities, etc.). In this respect particularly promising is the search for the chemical signature[20] of the mass-transfer process for the BSSs along the red sequence, even though these observations represent a real challenge for the current generation of high resolution spectrographs mounted at 8-10m class telescopes.

1. Marín-Franch, A. *et al.* The ACS Survey of Galactic Globular Clusters. VII. Relative Ages. *Astrophys. J.* **694**, 1498-1516 (2009)

2. Shara, M. M., Saffer, R. A., & Livio, M. The First Direct Measurement of the Mass of a Blue Straggler in the Core of a Globular Cluster: BSS19 in 47 Tucanae *Astrophys. J. Lett.* **489**, L59-L63 (1997)

3. McCrea, W. H. Extended main-sequence of some stellar clusters. *Mon. Not. R. Astron. Soc.* **128**, 147-155 (1964)

4. Hills J. G., & Day, C. A. Stellar Collisions in Globular Clusters. *Astrophys. J. Lett.* **17**, 87 (1976)

5. Knigge, C., Leigh, N. & Sills, A. A binary origin for `blue stragglers' in globular clusters, *Nature*, **457,** 288-290 (2009)

6. Meylan, G. & Heggie, D.C., Internal dynamics of globular clusters. *Ann. Rev Astron. & Astrophys.* **8**, 1-143 (1997)

7. Ferraro, F.R. & Lanzoni, B. Blue Straggler Stars in Galactic Globular Clusters: Tracing the Effect of Dynamics on Stellar Evolution. In *Dynamical Evolution of Dense Stellar Systems*, Proceedings of the International Astronomical Union, IAU Symposium, **246**, 281-290 (2008)

8. Mapelli, M., Sigurdsson, S., Ferraro, F. R., Colpi, M., Possenti, A. & Lanzoni, B. The radial distribution of blue straggler stars and the nature of their progenitors. *Mon. Not. R. Astron. Soc.* **373**, 361-368 (2006)

9. Trager, S. C., Djorgovski, S. & King, I.R., Structural parameters of Galactic Globular Clusters in *Structure and Dynamics of Globular Clusters,* Editors, S.G. Djorgovski and G. Meylan; Publisher, *Astronomical Society of the Pacific Conference Series*, **Vol. 50**, 347-355 (1993)

10. Bailyn, C. D. Are there two kinds of blue stragglers in globular clusters? *Astrophys. J.* **392**, 519-521 (1992)

11. Leonard, P.J.T. Stellar collisions in globular clusters and the blue straggler problem. *Astronom. J.*, **98**, 217-226 (1989)

12. Ferraro, F. R., Sills, A., Rood, R.T., Paltrinieri, B. & Buonanno, R. Blue Straggler Stars: A Direct Comparison of Star Counts and Population Ratios in Six Galactic Globular Clusters. *Astrophys. J.* **588**, 464-477 (2003)

13. Sills, A., Karakas, A.I. & Lattanzio, J. Blue Stragglers After the Main Sequence. *Astrophys. J.* **692**, 1411-1420 (2009)

14. Davies, M.B., Piotto, G. & de Angeli, F., Blue straggler production in globular clusters *Mon. Not. R. Astron. Soc.* **349**, 129-134 (2004)

15. Cariulo, P., Degl'Innocenti, S. & Castellani,V. Calibrated stellar models for metal-poor populations. *Astron. & Astrophys.* **421**, 1121-1130 (2004)

16. Ferraro, F.R., Messineo, M., Fusi Pecci, F., De Palo,M.A., Straniero, O., Chieffi, A. & Limongi, M. The Giant, Horizontal, and Asymptotic Branches of Galactic Globular Clusters. I. The Catalog, Photometric Observables, and Features. *Astronom. J.*, **118**, 1738-1758 (1999)

17. Tian, B., Deng, L. , Han, Z. & Zhang, X.B. The blue stragglers formed via mass transfer in old open clusters. *Astron. & Astrophys.* **455**, 247-254 (2006)

18. McMillian, S., Hut, P. & Makino, J. Star cluster evolution with primordial binaries. I - A comparative study. *Astrophys. J.* **362**, 522-537 (1990)

19. Hurley, J., et al. Deep Advanced Camera for Surveys Imaging in the Globular Cluster NGC 6397: Dynamical Models. *Astronom. J.*, **135**, 2129-2140 (2008)

20. Ferraro, F.R., et al. Discovery of Carbon/Oxygen-depleted Blue Straggler Stars in 47 Tucanae: The Chemical Signature of a Mass Transfer Formation Process. *Astrophys. J. Lett.* **647**, L53-L56 (2006)

21. Stetson, P.B. DAOPHOT - A computer program for crowded-field stellar photometry. *PASP*, **99**, 191-222 (1987)

22. Stetson, P.B. The centre of the core-cusp globular cluster M15: CFHT and HST Observations, ALLFRAME reductions. *PASP*, **106**, 250-280 (1994)

23. Pietrukowicz, P., & Kaluzny, J. Variable Stars in the Archival HST Data of Globular Clusters M13, M30 and NGC 6712. *Acta Astronomica*, **54**, 19-31 (2004)

24. Vilhu, O. Detached to contact scenario for the origin of W UMa stars. *A&A*, **109**, 17-22 (1982)

25. Hartigan P., Applied Statistics, Vol. 34, N. 3, p. 320 (1985)

26. King, I.R. The structure of star clusters. III. Some simple dynamical models. *Astronom. J.* **71**, 64-75 (1966)

27. Harris, W.E. A Catalog of Parameters for Globular Clusters in the Milky Way. *Astronom. J.*, **112**, 1487-1488 (1996)

28. Noyola, E. & Gebhart, K. Surface Brightness Profiles for a Sample of LMC, SMC, and Fornax Galaxy Globular Cluster. *Astronom. J.*, **132**, 447-466 (2006)

29. Djorgovski, S. Physical parameters of Galactic Globular Clusters in *Structure and Dynamics of Globular Clusters*, Editors, S.G. Djorgovski and G. Meylan; Publisher, *Astronomical Society of the Pacific Conference Series*, **Vol. 50**, 373-382 (1993)

**Supplementary Information** accompanies the paper on **www.nature.com/nature**.

**Acknowledgements:** This research was supported by "Progetti di Ricerca di Interesse Nazionale 2008" granted by the Istituto Nazionale di Astrofisica. We acknowledge the financial support of the Agenzia Spaziale Italiana and the Ministero dell'Istruzione, dell'Universitá e della Ricerca. FRF, BL, ED and AS thank the "Formation and evolution of Globular Clusters programme" and the Kavli Institute for Theoretical Physics in Santa Barbara (California, USA) for the hospitality during their stay, when the motivations of this project were discussed and the work planned. We acknowledge support from the ESTEC Faculty Visiting Scientist Programme. RTR is partially supported by a STScI grant. This research has made use of the ESO/ST-ECF Science Archive facility which is a joint collaboration of the European Southern Observatory and the Space Telescope - European Coordinating Facility.

**Author contributions:** F.R.F. designed the study and coordinated the activity. G.B., E.D. and S.B. analysed the data. A.S. and A.I.K. developed collisional models. B.L. and P.M. computed the surface density profile and performed comparisons with a single mass King model. F.R.F. and B.L. wrote the paper. F.FP., A.S. and R.T.R. critically contributed to the paper discussion and presentation. All the authors contributed to discuss the results and commented on the manuscript.

Correspondence to: F.R.Ferraro[1] Correspondence and requests for materials should be addressed to F.R.F.
**francesco.ferraro3@unibo.it**

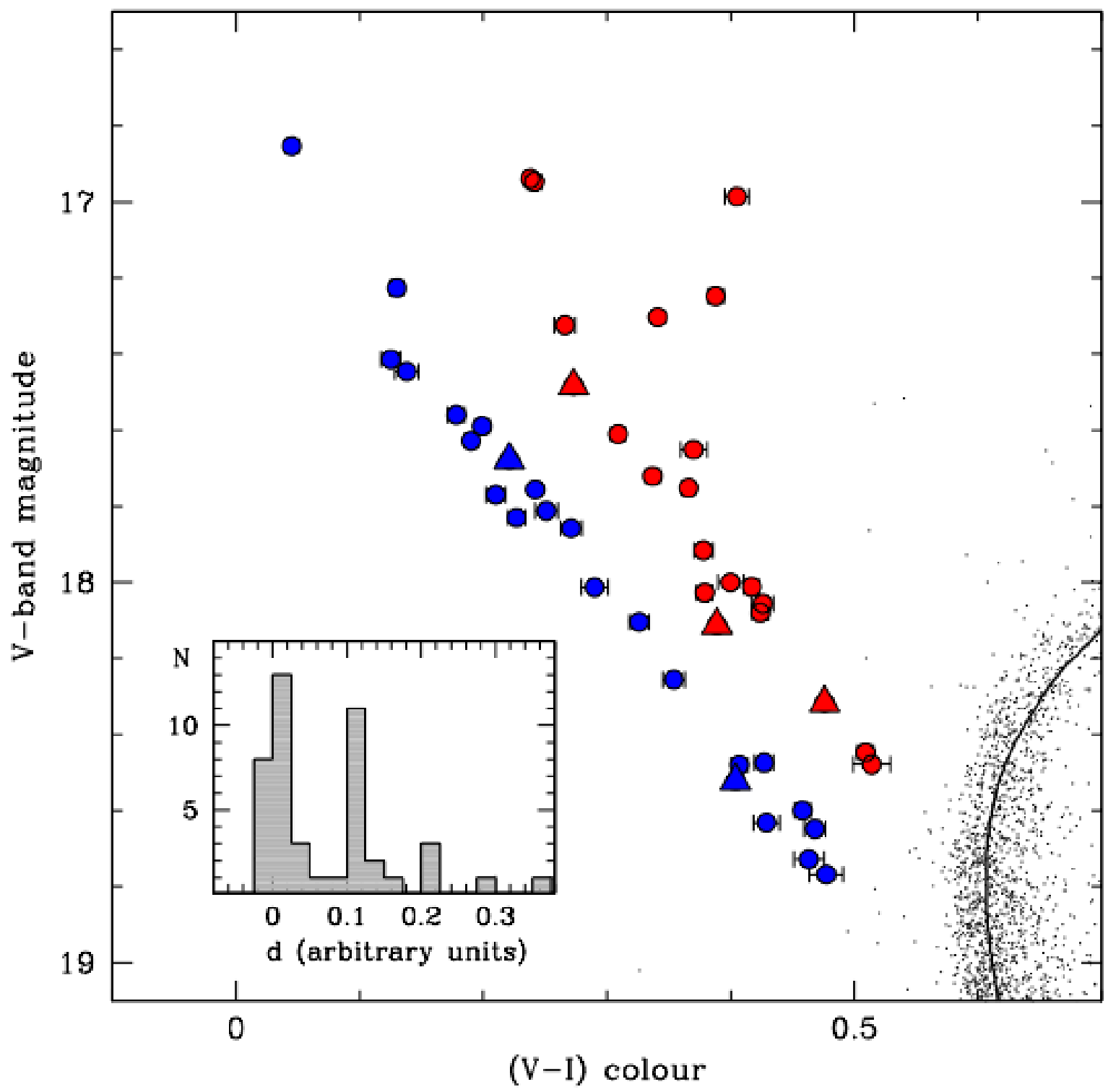

**Figure 1. The two blue straggler sequences of M 30.** BSS region of the (V,V-I) CMD. The selected BSSs are plotted as large circles, with the red and blue colours distinguishing the red and the blue BSS sequences, respectively. A series of 44 images (22 in each filter), secured with the Wide Field Planetary Camera 2 (WFPC2) through the F814W (I band) and the F555W (V band) filters, has been analyzed by using PSF-fitting photometry[21,22]. Errors (1 s.e.m) in magnitude and colours have been computed from repeated measures and are also plotted (they are typically lower than 0.01 mag; in most cases the error bars are smaller than the point size). The two sequences are separated in magnitude by $\Delta$V ~ 0.4 mag and in colour by $\Delta$ (V-I) ~ 0.12 mag.

Using these time-series, we have tested the variability of the selected BSSs and found five candidate variables (triangles): on the basis of the light curve characteristics, the three brightest variables have been classified[23] as a W Ursae Majoris (W Uma) contact binaries. The two faintest candidates show quite scattered light curves that prevent a reliable classification. W Uma stars are binary systems losing orbital momentum because of magnetic braking. These shrinking binary systems, which are initially detached, evolve to the contact stage and finally merge into a single star. The evolution of W UMa systems is thought to be a viable channel for the formation of BSS[20,24]. Inset, distribution of the geometrical distances, d, of the selected BSSs from the straight line that best-fits the blue BSS sequence. Two well-defined peaks are clearly visible. A *Dip-test*[25] applied to this distribution demonstrates that it is bimodal at a significance level of more than $4\sigma$, confirming that the two sequences are nearly parallel to each other.

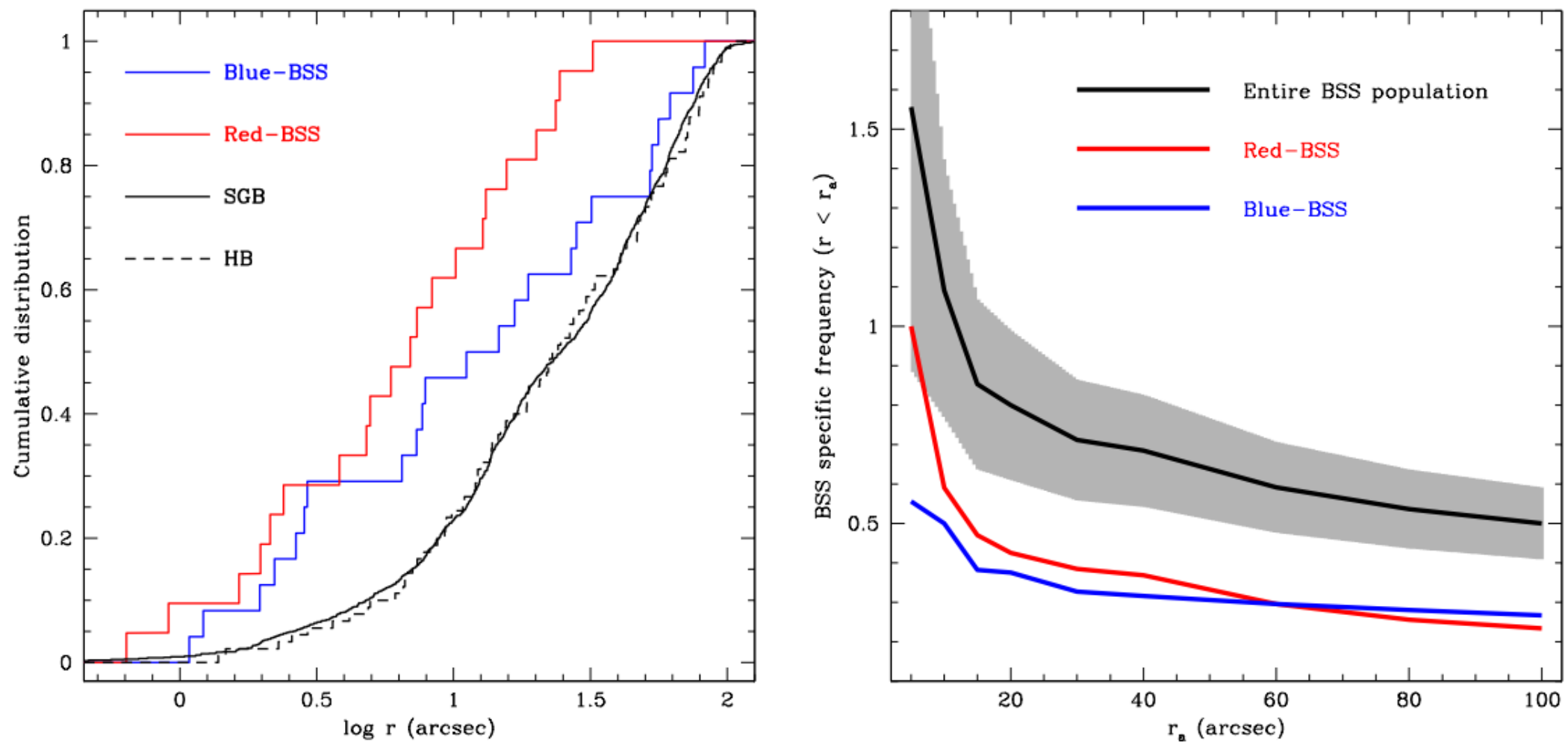

**Figure 2. The BSS radial distribution.**

**a,** Cumulative radial distribution of red BSSs (red line) and blue BSSs (blue line), as functions of the projected distance, r, from the cluster centre of gravity. The distribution of subgiant-branch stars (solid black line) and horizontal-branch stars (dashed black line) is also plotted for comparison.

**b,** BSS specific frequency computed in circular areas of increasing radius $r_a$. The lines correspond to the overall BSS population (black), and to the red BSS (red) and blue BSS (blue) subpopulations separately. The grey area around the black line shows the 1σ s.d. uncertainty in the specific frequency. BSSs are substantially more numerous than HB stars in the cluster centre. Although the small number of stars in the sample prevents statistical robustness in our results, we note that in the innermost 5-6 arcsec (~0.2 pc) the red BSSs tend to be as numerous as the horizontal branch stars and dominate the ratio.

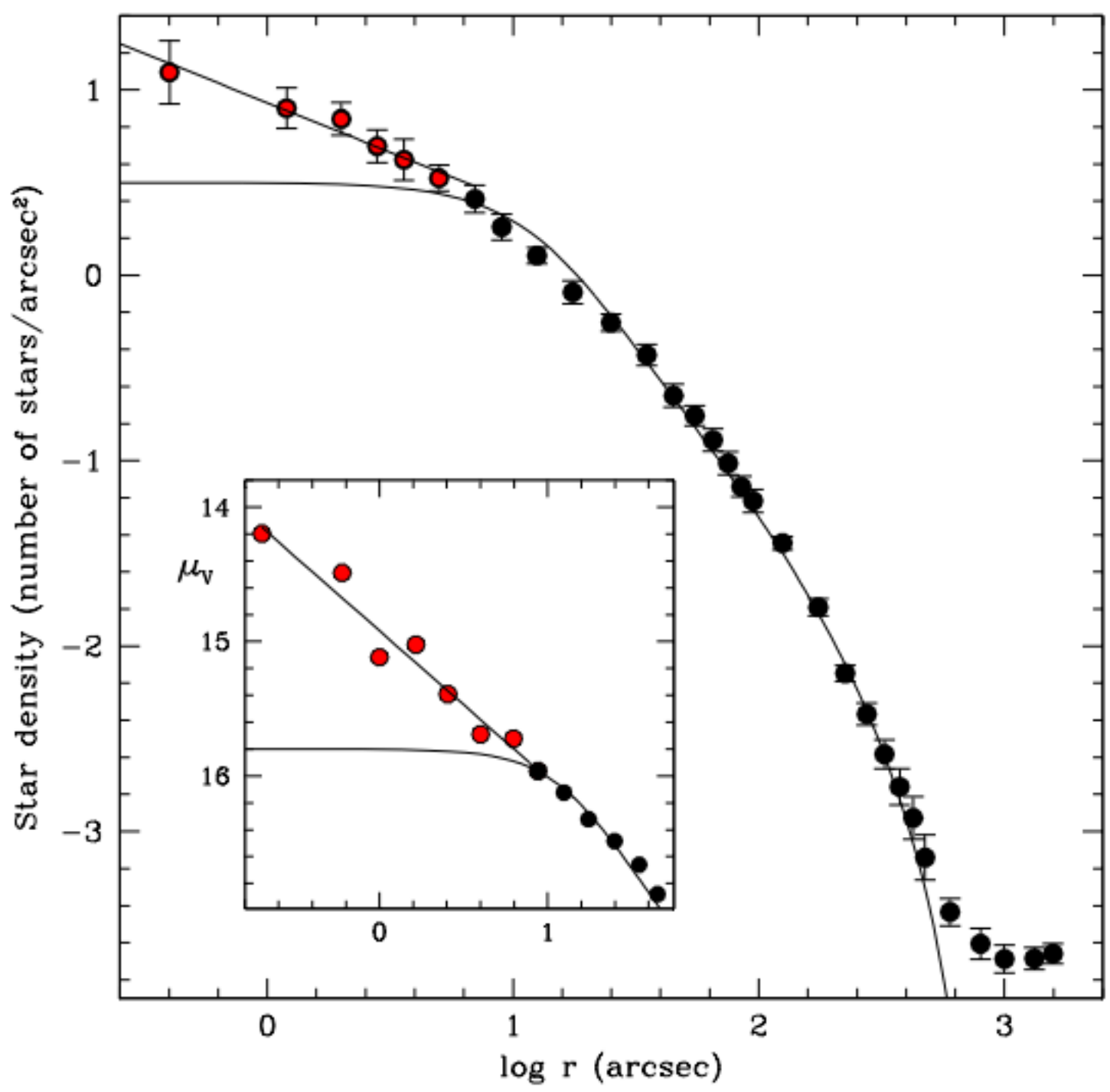

**Figure 3. The star density profile of M 30.** Profile (errors, 1 s.e.m) obtained from resolved star counts over the entire cluster extension: the WFPC2 data set was combined with ACS data and with wide-field ground-based observations secured at the European Southern Observatory New Technology Telescope (La Sill Observatory, Chile) and the Canada-France-Hawaii Telescope MegaCam (Hawaii). The single-mass King model[26] that best fits the observed profile excluding the innermost ($r$ < 5 arcsec) points is shown as a thick solid line. The points that deviate from the King profile (red) are well fitted by a power-law with slope $\alpha \approx -0.5$ (thin solid line). Inset, surface brightness profile derived from the WFPC2 V-band images within the innermost 40 arcsec, with the two lines having the same meaning as above. The measured central surface brightness is $\mu_V \sim 14.2$ mag arcsec$^{-2}$. This value is significantly lower (corresponding to greater brightness) than that listed in currently adopted cluster catalogues[27], but fully consistent with that obtained in most recent studies[28]. Following the procedure described in the literature[29] and adopting a distance of 8.75 kpc and a reddening E(B-V) = 0.03[16], we derived $\nu \sim 9.6 \times 10^4$ $L_{\odot}$ $pc^{-3}$ for the luminosity density within the density cusp (that is, for *r* < 5 arcsec; $L_{\odot}$, solar luminosity). Under the assumption of a mass-to-light ratio of three and a mean stellar mass *of* 0.5 $M_{\odot}$, this corresponds to a number density of stars $n \sim 5.8 \times 10^5$ $pc^{-3}$ ( Supplementary Table 1; $M_{\odot}$ solar mass).

Both profiles were computed with respect to the newly determined cluster centre of gravity (Supplementary Table 1), which is located at ~3 arcsec southeast of the centre listed in commonly used catalogues of globular-cluster parameters[27].

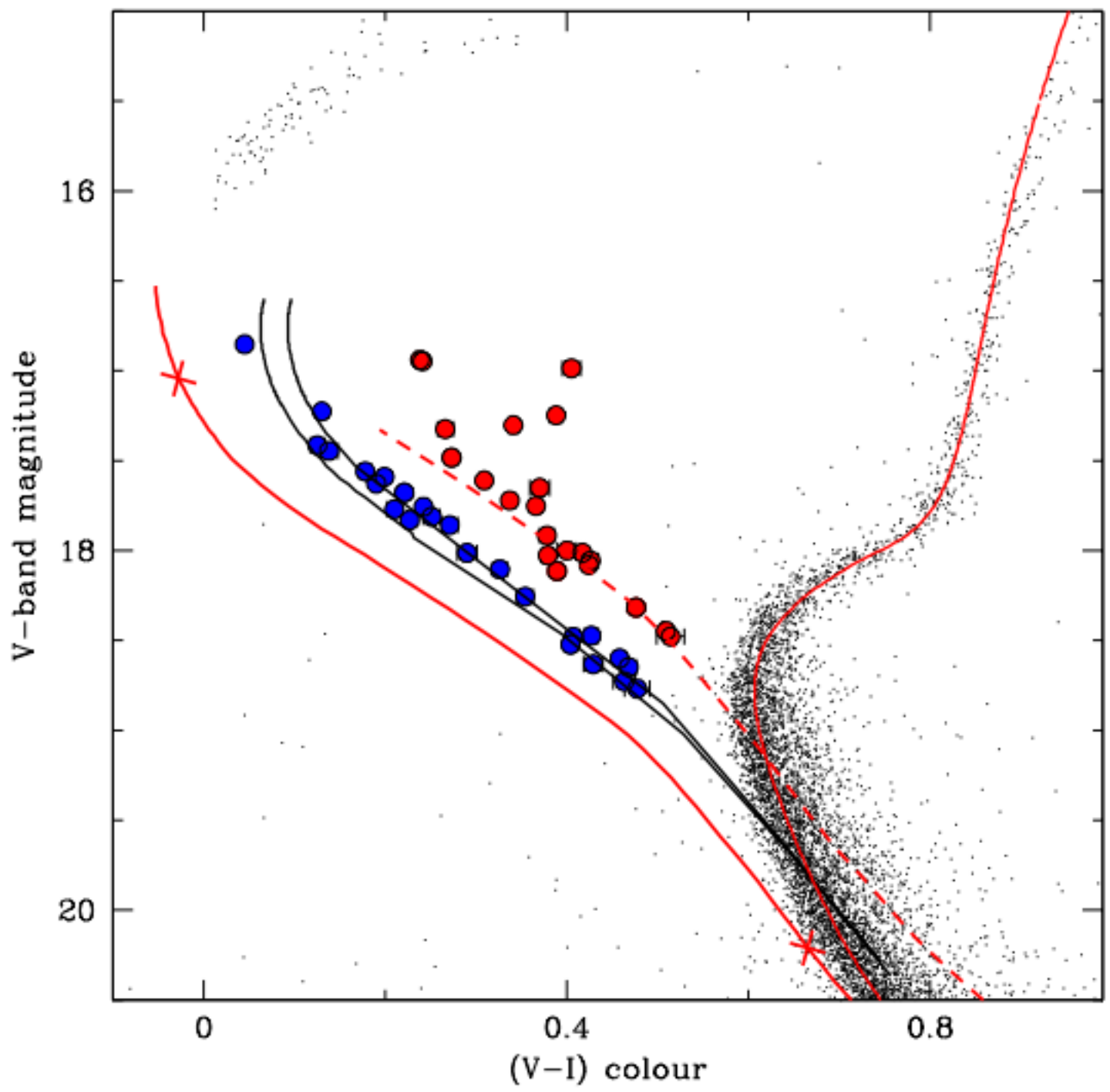

**Figure 4. Comparison with collisional and binary evolution models.** Magnified portion of the CMD of M 30. The solid black lines correspond to the collisional isochrones[13] corresponding to 1 and 2 Gyr, respectively, which accurately reproduce the blue BSS sequence. The red solid lines correspond to the single-star isochrones[15] respectively corresponding to 13 Gyr (well fitting the main cluster evolutionary sequences) and to 0.5 Gyr (representing the reference cluster zero-age main sequence (ZAMS)). The two crosses mark the respective position of a 0.8 star and a 1.6 $M_{\odot}$ star along the ZAMS. The dashed red line corresponds to the ZAMS shifted by 0.75 mag, marking the position of the 'low-luminosity boundary' predicted[17] for a population of mass-transfer binary systems. This line well reproduces the red BSS sequence.

# Supplementary Information

**Data-set:** The data-set used in this study consists of a series of 44 images (22 in filter F814W and 22 in filter F555W) secured in 1999 with the Wide Field Planetary Camera 2 (WFPC2) on board the Hubble Space Telescope. Exposures have been obtained within a project (GO-7379, PI: Edmonds) aimed at searching for stellar variability, from main sequence binaries, cataclysmic variables and blue stragglers, in the congested core of M30[23,30].

**Core collapse:** The core collapse is a catastrophic dynamical process consisting in the runaway contraction of the core of a star cluster[6]. Binary-binary and binary-single collisions are thought to halt (or delay) the collapse of the core, thus avoiding infinite central densities[6]. A common core-collapse observational signature is a steep cusp in the projected star density profile. About 15% of the globular cluster population in our Galaxy (including M30) shows evidence of this phenomenon[9].

# Supplementary Table 1

**Table 1 Cluster parameters**

| | |
|---|---|
| Centre of gravity | $\alpha$=21$^h$40$^m$22.13$^s$ $\delta$=-23°10'47.4" |
| Central surface brightness | $\mu_V$ = 14.2 mag arcsec$^{-2}$ |
| Physical size of the central cusp | $r_{cusp}$ = 0.2 pc |
| Central mass density [$M_\odot$ pc$^{-3}$] | Log $\rho_0$ = 5.48 |
| True distance modulus[16] | $(m-M)_0$ = 14.71 mag |
| Colour excess[16] | E(B-V) = 0.03 mag |
| Distance[16] | d = 8.75 kpc |
| Integrated V magnitude[27] | V=7.19 mag |
| Integrated absolute magnitude | $M_V$ = -7.61 mag |
| Age | t = 13 Gyr |
| Metallicity[16] | [Fe/H] = -1.9 |

# Supplementary References

30. Lugger, P. M.; Cohn, H. N., Heinke, C. O., Grindlay, J. E. & Edmonds, P.D. Variable Chandra X-Ray Sources in the Collapsed-Core Globular Cluster M30 (NGC 7099). *Astrophys. J.* **657**, 286-301 (2007)